\documentclass{amsart}
\usepackage{amssymb, amsmath, amsthm}
\usepackage{physics}
\usepackage{mathrsfs}
\usepackage{amscd}
\usepackage[active]{srcltx}
\usepackage{verbatim}
\usepackage{graphicx}
\usepackage{subcaption}
\usepackage{float}

\usepackage[colorlinks,linkcolor={blue}, citecolor={blue},urlcolor={red},]{hyperref}

\let\mathcal \undefined
\def\mathcal{\mathscr}

\theoremstyle{plain}

\theoremstyle{remark}

\theoremstyle{plain}

\numberwithin{equation}{section}

\newcommand{\up}{\uparrow}
\newcommand{\down}{\downarrow}

\newcommand{\C}{\mathbb{C}}

\allowdisplaybreaks

\begin{document}

\title[Delayed choice]{Delayed choice experiments: \\ An Analysis in forward time}

\author{Marijn Waaijer \& Jan van Neerven}

\address{Delft Institute of Applied Mathematics\\
Delft University of Technology\\P.O. Box 5031\\2600 GA Delft\\The Netherlands}
\email{waaijermarijn@gmail.com, j.m.a.m.vanneerven@tudelft.nl}

\date{\today}

\begin{abstract}
In this article, we present a detailed analysis of two famous delayed choice experiments: Wheeler's classic gedanken-experiment and the delayed quantum eraser. {Our analysis} shows that the outcomes of both experiments can be fully explained on the basis of the information collected during the experiments using textbook quantum mechanics only. At no point in the {argument,}
information from the future is needed to explain what happens next. In fact, more is true: for both experiments, we show, in a strictly mathematical way, that a modified version in which the time-ordering of the steps is changed to avoid the delayed choice leads to exactly the same final state. In this operational sense, the scenarios are completely equivalent in terms of conclusions that can be drawn from their outcomes.
\end{abstract}


\maketitle

\section{Introduction}

Delayed choice experiments constitute a class of experiments with the general feature that ``quantum effects can mimic an influence of future actions on past events" \cite[Conclusion and outlook]{Ma}. The keyword in this characterisation is `can mimic'. As the following remark by Ma et al. \cite{Ma2} in their article on delayed-choice entanglement swapping\footnote{We will not address the delayed choice entanglement swapping experiment in the remainder of this article. It can be analysed in the same way as done here for the Wheeler's gedanken-experiment and the quantum erasure experiment, and this analysis leads to similar conclusions.} perfectly highlights, the status of the retro-causation of delayed choice experiments is generally understood to be dependent on the interpretation of quantum mechanics one adheres to:

\begin{quote}If one viewed the quantum state as a real physical object, one could get the paradoxical situation that future actions seem to have an influence on past and already irrevocably recorded events. However, there is never a paradox if the quantum state is viewed as no more than a `catalogue of our knowledge'. Then the state is a probability list for all possible measurement outcomes, the relative temporal order of the three observers' events is irrelevant and no physical interactions whatsoever between these events, especially into the past, are necessary to explain the delayed-choice entanglement swapping.
\end{quote}

\noindent
Thus, according to Ma et al., whether delayed choice experiments show apparent retro-causality cannot be decided solely from their outcome data as such. It depends on the role we ascribe to the state. If the state is considered to correspond to an objective part of reality, then these experiments seem to show retro-causality; if, on the other hand, the state is considered merely a catalogue of our knowledge, then the feature of retro-causality disappears. \par

In this paper we challenge this conclusion. Our central claim is that the outcomes of delayed choice experiments can be fully explained in terms of a step-by-step mathematical analysis in forward time. This analysis involves no retro-causal steps, regardless of the ontological status one wishes to ascribe to the quantum mechanical state as such (real or epistemological). We substantiate our claim by providing such step-by-step analyses for two celebrated delayed choice experiments: Wheeler's original gedanken-experiment \cite{Wheeler, Wheeler2} and the `delayed quantum eraser' of Scully and Dr\"{u}hl \cite{ScullyDruehl}. For both experiments, we provide a parallel analysis of their delayed and non-delayed counterparts and show that both lead to the same final quantum state. This is neither `obvious' nor interpretation-dependent but represents the outcome of a careful analysis of the steps involved. It demonstrates that, on an operational level, the delayed and non-delayed versions of the experiments cannot be distinguished. As we will argue in the final section, the puzzling aspects of delayed choice experiments seem to arise from the use of notions such as `wave-particle duality' and `which path information' as explanatory rather than descriptive tools. As these notions make sense only once the completed experiment can be overseen in its entirety, their use entails a certain degree of backward-in-time reasoning about the experiments. By abandoning these descriptive notions in favour of a forward mathematical analysis, the puzzling aspects of the delayed choice experiments disappear.

There is extensive literature discussing theoretical aspects and experimental realisations of delayed choice experiments. A (quantum) Wheeler's delayed choice experiment, in which the second beam splitter can be in a superposition of being present and absent, was proposed about a decade ago by \cite{IonTer}; for a fairly recent review, the reader is referred to \cite{Ma}. Additional references {to some recent papers} can be found in, e.g., \cite{Qur}. General aspects of the problem of retro-causality in quantum mechanics are discussed in \cite{Faye, FE, Wharton, Wharton2}, {where further references can be found}. \par
The arguments presented here can be placed in line with some of these earlier studies. In her discussion of the delayed quantum eraser, Hossenfelder \cite{Hossenfelder} points out how the data of the first, signal particle remains necessarily unchanged during the whole experiment; the purported retro-causality of the delayed eraser therefore happens only on the level of retro-active selections in the signal data based on the actions of the second, idler photon. Our mathematical analysis confirms this position.
Kastner \cite{Kastner} argues that the delayed quantum eraser is in essence no different from a standard EPR-pair, in that the order of space-like separated operations is irrelevant; the quantum eraser does not `delay' nor `erase'.
Gaasbeek \cite{Gaasbeek} argues that, from the point of view of the theory of special relativity, the sequence of the space-like separated measurements is relative to the observer overseeing the experiment, and therefore the order between these measurement operations should have no influence on the eventual outcomes. We agree with Kastner and Gaasbeek. However,
our point is not made by analogy to an EPR-pair, but through direct analysis of the delayed quantum eraser itself, and
we arrive at our conclusion by showing thus mathematically using standard textbook quantum mechanics only; our arguments do not invoke special relativity. 
In \cite{DonRaeMic}, {Donker et al.} argue that the outcomes of delayed choice experiments can be explained entirely in terms event-by-event based models involving  objects travelling one-by-one through the experimental set-up and generating clicks of a detector.
Let us finally mention the work \cite{DGPAS} by Dieguez et al., which analyses delayed choice experiments by adopting an operational quantifier of realism, enabling them argue that the visibility at the output has no connection whatsoever with wave and particle elements of reality as defined in accordance with the adopted criterion of realism. In the same paper it is observed that ``{\em To date, a detailed analysis is lacking which would
allow one to track the behavior of the system at every stage of the
experiment''}. We provide a framework for such an analysis here.

\section{Wheeler's Delayed Choice experiment \label{sec:Wheelersexperiment total}}

\begin{figure*}
\begin{subfigure}{0.3\textwidth}
 \includegraphics[scale = 0.45]{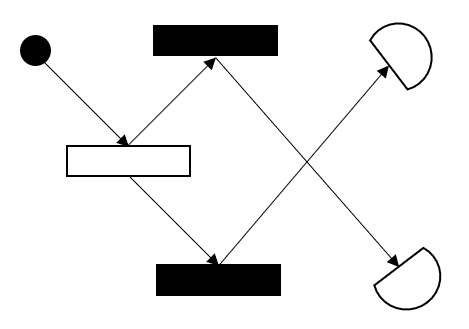}
\end{subfigure}
\hskip1.5cm
\begin{subfigure}{0.35\textwidth}
 \includegraphics[scale = 0.45]{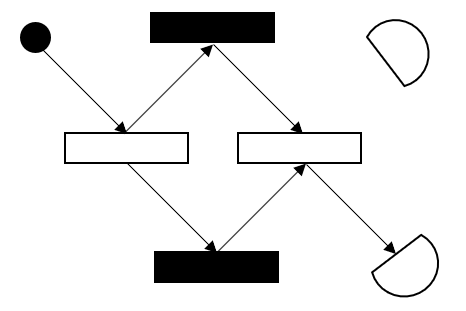}
\end{subfigure}
 \caption{\footnotesize The Mach-Zehnder Interferometer and without a second beamsplitter. The black dot denotes the single-photon source, the open rectangle denotes a beamsplitter, the closed rectangles are mirrors, and the open-half circles are the detectors. The arrows represent the possible `paths' of the photons emitted by the laser. The beamsplitter creates a superposition describing the two possible paths depicted in the figure. As explained in the main text, the second beamsplitter has the effect that no incoming photons are detected by the top detector due to destructive interference.\label{fig:MZN}}
\end{figure*}

Wheeler's delayed choice experiment \cite{Wheeler, Wheeler2} (as cited from \cite[sect. II.D]{Ma}) makes use of two Mach-Zehnder interferometers, experimental devices that can be used to demonstrate certain quantum mechanical phenomena involving superposition. A Mach-Zehnder interferometer consists of a single photon source and a sequence of mirrors and beamsplitters which eventually steer the emitted photon towards (one of) two detectors.
In the first set-up, only one beamsplitter is used. A single photon is emitted towards the beamsplitter, which brings the photon into a superposition of travelling via the upper or lower path (see Figure \ref{fig:MZN}, left). After being directed towards the detectors by mirrors, the photon is then detected in one of the detectors with equal probability.
The second set-up of the experiment follows the same description, but after having been deflected by the mirrors the photon encounters a second beamsplitter, which causes both paths to interfere. This causes only the bottom detector to detect incoming photons (see Figure \ref{fig:MZN}, right).

\begin{figure}[ht]
 \centering
 \includegraphics[scale = 0.45]{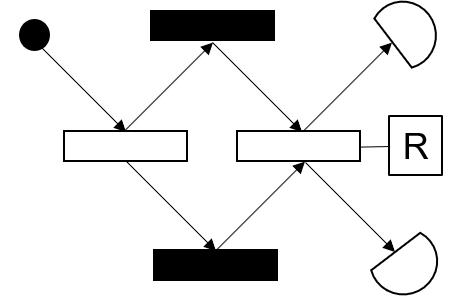}
 \caption{\footnotesize A delayed choice experiment with a Mach-Zehnder Interferometer. The square labeled $R$ denotes a quantum random bit generator which, depending on its two possible outputs, activates or deactivates the second beamsplitter.}
 \label{fig:MZDelayedChoice}
\end{figure}

In 1978, Wheeler
proposed to combine the two scenarios by having a quantum random bit generator decide between them.
If the quantum random generator gives output `off', the second beamsplitter is deactivated and the first scenario is followed;
if the quantum random generator gives output `on', the second beamsplitter is activated and the second scenario is followed (see Figure \ref{fig:MZDelayedChoice}). This set-up becomes a `delayed-choice' experiment by letting the quantum random generator decide between the two scenarios only {\em after} the photon has passed the first beamsplitter. {Experimental realisation of this gedanken-experiment was reported in \cite{JGTGAR}.}\par

\subsection{Wheeler's original argument\label{sec:wheelersoriginalexperiment}}

Wheeler's original interest in the above experiment came from an argument based on the Copenhagen interpretation. In this view, the separate clicks of the two detectors in the first scenario are interpreted as revealing the `particle' nature of the photon, whereas the presumed interference explaining the second scenario is interpreted as revealing the `wave' nature of the photon. From this point of view, the introduction of the second beamsplitter ``forces'' the photon to behave like a particle {\em or} like a wave. In the delayed choice experiment, this choice can only be made ``in flight'' once it is known whether or not a second beamsplitter will be encountered. Thus Wheeler interprets this experiment as a manifestation
of retro-causation:

\begin{quote}``In this sense, we have a strange inversion of the normal order of time.
We, now, by moving the mirror in or out have an unavoidable effect on what we have a right to say about the already past history of that photon.'' ($\dots$) ``Thus one decides whether the photon `shall have come by one route [as particle] or by both routes [as an interfering wave]’ after it has already done its travel'' \cite{Wheeler} (cited directly from \cite{Ma}). \end{quote}

\subsection{Analysis of the experiment} \label{sec:analysiswheeler} In this section, we provide our analysis of Wheeler's gedanken-experiment. We first provide an analysis of the experiment in the language of standard quantum mechanics and, subsequently, recast it in mathematical language. We use the latter to show the equivalence of both the delayed and non-delayed versions of the experiment. Lastly, we ground this equivalence in the physical principle that space-like separated arms of the experiment necessarily commute.

\subsubsection{Quantum mechanical description of the gedanken-experiment}\label{sec:timeanalysiswheeler}
Let us first analyse the scenario without a quantum random bit generator as depicted on the left in Figure \ref{fig:MZN}.
In this scenario, no second beamsplitter is present. The states of the photon before passing the beamsplitter, after passing the
beamsplitter and before arriving at the mirrors, and after being deflected by the mirrors but before arriving at the detectors,
can be described as follows:
\begin{equation}\label{delayedchoiceexp3-1}
\begin{aligned}
 \ket{1, t = 1} & = \ket{\down},\\
 \ket{1, t = 2} & = \frac{1}{\sqrt{2}}(\ket{\up} - \ket{\down}), \\
 \ket{1,t = 3} & = \frac{1}{\sqrt{2}}(\ket{\down} - \ket{\up}).
\end{aligned}
\end{equation}
The number `1' refers to this first scenario. \par
In the second scenario, with a second beamsplitter inserted between the mirrors and the detectors as depicted on the right in Figure \ref{fig:MZN}, the first three stages are the same, but we must add a fourth stage describing the state of the photon after it has encountered the second beamsplitter. The second beamsplitter
changes the state $\ket{\up}$ to $\frac{1}{\sqrt{2}} (\ket{\down}+\ket{\up})$ and $\ket{\down}$ to $\frac{1}{\sqrt{2}} (\ket{\up}-\ket{\down})$.
Inserting this into the third line of Equation \eqref{delayedchoiceexp3-1}, we arrive at
\begin{align*}
 \ket{2, t = 4}
 & = \frac{1}{\sqrt{2}}\Biggl(\Bigl(\frac{1}{\sqrt{2}} (\ket{\up}-\ket{\down})\Bigr) -\Bigl(\frac{1}{\sqrt{2}} (\ket{\down}+\ket{\up}) \Bigr)\Biggr)
 =-\ket{\down}.
\end{align*}
In the third scenario (as in Figure \ref{fig:MZDelayedChoice}), a quantum random bit generator chooses between these two scenarios {\em after a photon has been deflected by the mirrors.} We model this by
coupling the photon to the quantum random bit generator, whose output we represent by the `off' state $\ket{\text{off}}$ (no second beamsplitter is introduced) and the `on' state $\ket{\text{on}}$ (the second beamsplitter is introduced). This results in the states
\begin{equation}\label{delayedchoiceexp3-1a}
\begin{aligned}
 \ket{3, t = 1} & = \ket{\down}\ket{\text{off}},\\
 \ket{3, t = 2} & = \frac{1}{\sqrt{2}}(\ket{\up} - \ket{\down})\ket{\text{off}}, \\
 \ket{3,t = 3} & = \frac{1}{\sqrt{2}}(\ket{\down} - \ket{\up})\ket{\text{off}},\\
 \ket{3, t = 4} & = \frac{1}{2} (\ket{\down}-\ket{\up})\ket{\text{off}} - \frac{1}{\sqrt{2}}\ket{\down}\ket{\text{on}}.
\end{aligned}
\end{equation}
In the last step, the results of the first and second scenarios are realised with equal probability.

\subsubsection{Mathematical formulation and equivalence with the non-delayed experiment.}\label{sec:wheelermathequivalence}

We now turn to casting the analysis above in the language of quantum operators.
For this { purpose} we use the following set-up and notation. We introduce the Hilbert spaces $H_{\rm ph} = \mathbb{C}^2$ and $H_{\rm r} = \mathbb{C}^2$ modelling the photon and the quantum random bit generator. The composite system of photon and random bit is modelled on the Hilbert space $H_{\rm ph}\otimes H_{\rm r}\eqsim \C^4$, which we think of as being endowed with the orthonormal basis
$$
 \left\{\ket{\up}\ket{\rm off} = \begin{pmatrix} 1 \\ 0 \\ 0 \\ 0\end{pmatrix}, \
 \ket{\up}\ket{\rm on} = \begin{pmatrix} 0 \\ 1 \\ 0 \\ 0 \end{pmatrix}, \
 \ket{\down}\ket{\rm off} = \begin{pmatrix} 0 \\ 0 \\ 1 \\ 0 \end{pmatrix}, \
 \ket{\down}\ket{\rm on} = \begin{pmatrix} 0 \\ 0 \\ 0 \\ 1 \end{pmatrix}
 \right\}.
$$
As before, the states $\ket{\rm on}$ and $\ket{\rm off}$ of the quantum random bit generator correspond to the presence, respectively absence, of the second beamsplitter. We model the various steps of the delayed choice experiment as unitary operators acting on $H_{\rm ph}\otimes H_{\rm r}$. With respect to the above basis, the first beamsplitter and the mirrors act on this tensor product respectively as $H\otimes I$ and $X\otimes I$, where the Hadamard operator $H$ and the `$X$-gate' operator $X$ act on $H_{\rm ph}$ by the unitary matrices
$$H =
\frac{1}{\sqrt{2}} \begin{pmatrix}
1 & \phantom{-} 1 \\ 1 & -1
\end{pmatrix},\qquad
X = \begin{pmatrix} 0 & 1 \\ 1 & 0 \end{pmatrix},
$$
respectively. The random number generator can be modelled as $I\otimes H$, with the Hadamard matrix acting on $H_{\rm r}$. Lastly, the dependence of the second beamsplitter on the state of the random generator can be modelled by the controlled Hadamard operator $R$ on $H_{\rm ph}\otimes H_{\rm r}$ given by

\begin{align*}
R =
\begin{pmatrix} 1 & 0 & 0 & 0 \\
 0 & \frac1{\sqrt 2} & 0 & \phantom{-}\frac1{\sqrt 2} \\
 0 & 0 & 1 & 0 \\
 0 & \frac1{\sqrt 2} & 0 &-\frac1{\sqrt 2} \\
 \end{pmatrix}.
\end{align*}

The experiment can then be described by following the sequence of operations on the original state to arrive at the final (to be measured) state. The photon first encounters a beamsplitter ($H\otimes I$), then a mirror ($X\otimes I$), after which the random generator is initiated ($I\otimes H$) and, lastly the photon encounters the second beamsplitter controlled by the random generator (R). Therefore the complete experiment can be described by

\begin{align}
\label{eq:wheeleroperationswithdelay}
A & := R \circ (I\otimes H) \circ (X\otimes I) \circ (H\otimes I).
\end{align}

Next we show the equivalence of the delayed experiment to the non-delayed experiment. In this case we operate the quantum random bit generator first, that is, after the photon has left the laser but before it arrives at the beamsplitter. In line with { the explanations provided in} the previous paragraph, this scenario can be modelled as
\begin{align} \label{eq:wheeleroperationsnodelay}
 A' & := R\circ (X\otimes I) \circ (H\otimes I) \circ (I\otimes H).
\end{align}
Then by direct computation of both operators (or by noting that the matrices $(X\otimes I) \circ (H\otimes I)$ and $I\otimes H$ commute) we see that
$$ A = \frac12 \begin{pmatrix}
1 & \phantom{-}1 & -1 & -1 \\
{\sqrt 2} & -{\sqrt 2} & \phantom{-}0 & \phantom{-}0 \\
1 & \phantom{-}1 & \phantom{-}1 & \phantom{-}1 \\
0 & \phantom{-}0 & -{\sqrt 2} & \phantom{-}{\sqrt 2} \\ \end{pmatrix} = A' .$$
From this result, we conclude that the delayed and non-delayed experiments are equivalent with respect to their mathematical description { in the sense that, as far as the final state is concerned,} it is irrelevant at which moment we operate the quantum random generator.

\subsubsection{The operational equivalence of the delayed and non-delayed experiment in relation to special relativity.} 
\begin{figure}[H]
 \centering
 \includegraphics[width = \textwidth]{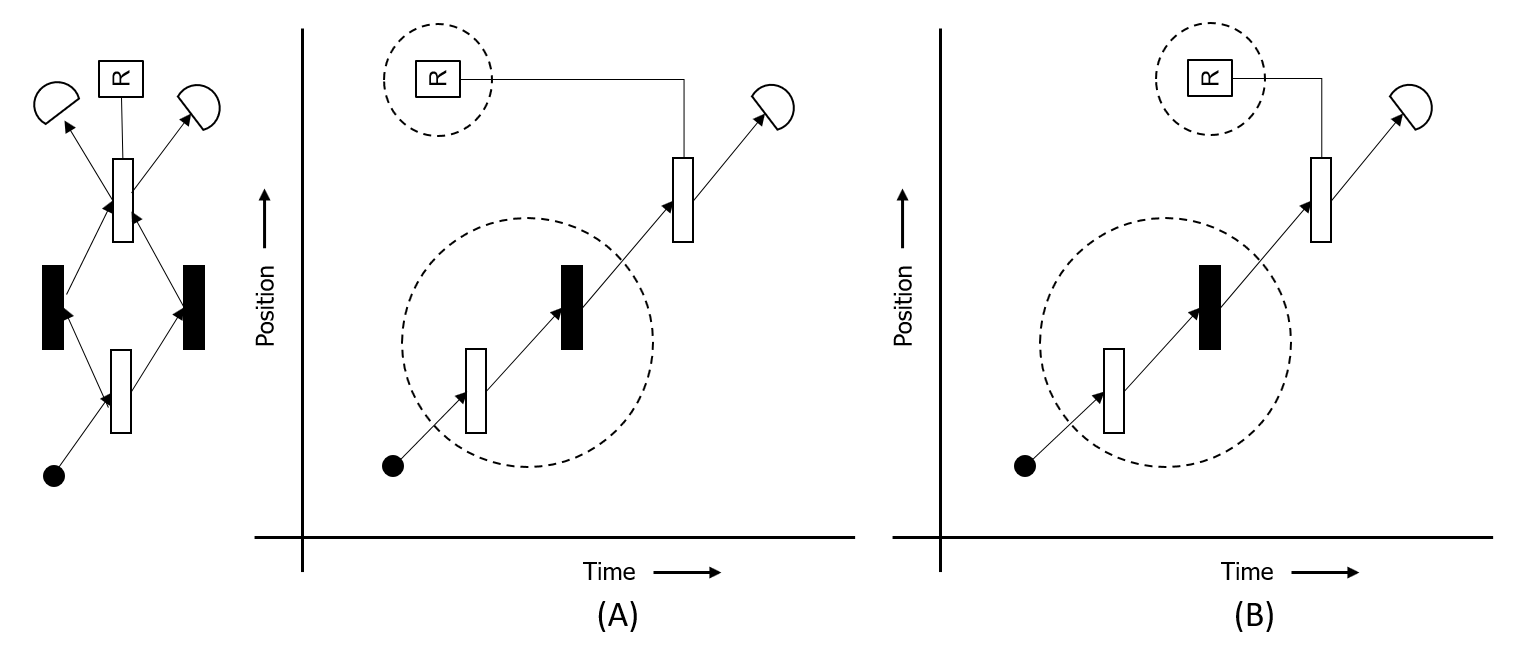}
 \caption{A schematic space-time diagram of Wheeler's delayed choice experiment with delay. As in Figure \ref{fig:MZDelayedChoice}, the black dot denotes the single-photon source, the open rectangle denotes a beamsplitter, the closed rectangles are mirrors, and the open-half circles denote the detectors. The two figures on the right display the temporal sequence of Wheelers experiment without delay in (A) and with delay in (B). The dotted lines in both figures indicate how the regions crucial for the change in temporal order are space-like separated. In the schematic, only the horizontal axis is included and therefore only one mirror and detector are depicted in the right-hand figures, but these represent both mirrors and both detectors.}
 \label{fig:wheeler_space_time_complete}
\end{figure}

\begin{figure}[H]
 \centering
 \includegraphics[width = \textwidth]{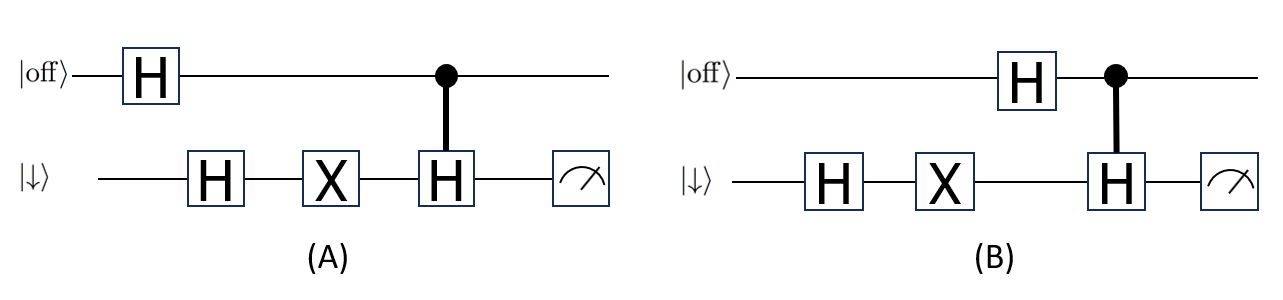}
 \caption{A quantum computing schematic of Wheeler's delayed choice experiment without delay in (A) and with delay in (B), matching the temporal sequences in figure \ref{fig:wheeler_space_time_complete}.}
 \label{fig:wheeler_quantum_computing_complete}
\end{figure}

Although our arguments do not depend on special relativity and are independent of certain parts of the experiment being space-time separated from others, it is of some interest to cast both the delayed and non-delayed formulations of Wheeler’s Gedanken-experiment into the space-time format presented in figure \ref{fig:wheeler_space_time_complete}. Not only do these figures isolate which operations are reversed between the delayed and non-delayed set-ups, but they also show that the equality is consistent with the relativistic principle that space-like separated operations should commute. Figure \ref{fig:wheeler_quantum_computing_complete} displays the descriptions of the experiments given in equations \eqref{eq:wheeleroperationswithdelay} and \eqref{eq:wheeleroperationsnodelay} in standard quantum computing language. These figures further clarify the equivalence $A = A'$.

\subsubsection{Conclusion}
Combining the arguments from the preceding sections, we conclude that the experimental set-up of Wheeler's gedanken-experiment with and without delayed are indistinguishable on the mathematical level and equivalent on the physical level. Therefore, from an operational point of view, in terms of predictions using standard quantum mechanics {\em no such thing as `delayed' choice exists} in this experiment.

\section{The delayed quantum eraser \label{sec:delayedquantumeraser}}

The most prevalent formulation of a delayed choice experiment today appears to be the `delayed quantum eraser' of Scully and Dr\"{u}hl \cite{ScullyDruehl}, experimentally realised by Kim et al. \cite{Kim}. In contrast to Wheeler's original idea, its formulation is not tied to any (Copenhagen-like) interpretation of the experiment. The delayed quantum eraser makes use of pairs of entangled particles. Its key feature is that after the first particle has been measured, it is possible to draw certain conclusions about it that seemingly depend on the outcome of the measurement of the second particle, which is performed at a later time. More concretely, upon repeating the experiment multiple times, we are able to identify subsets in our data of the first-measured particles which display either wave-like or particle-like behaviour, using the information obtained by measuring their entangled twins at a later moment. The puzzling thing is that wave-like or particle-like behaviour of the first-measured particles appeared to be already hidden in the data before we measured the entangled twins. As such, future events seem to retro-actively exert an influence on past events, and this retro-active influence even seems to affect already measured data. \par

\subsection{Set-up and results\label{sec:DQEsetupsandresults}}

The delayed quantum eraser experiment starts by sending a single photon towards a standard double slit, reminiscent of Young's famous double slit experiment.
Behind the slits, a nonlinear crystal converts the incoming photon into an entangled pair of photons of half the frequency. The first photon of this pair is called the {\em signal photon} and is sent towards a screen. The second photon is called the {\em idler photon} and is sent towards a set-up of (half-) mirrors and detectors in order to achieve the delayed erasure.
As was the case with the delayed choice experiment, this set-up is used to combine two experiments.

\begin{figure}
 \centering
 \includegraphics[width = 0.7\textwidth]{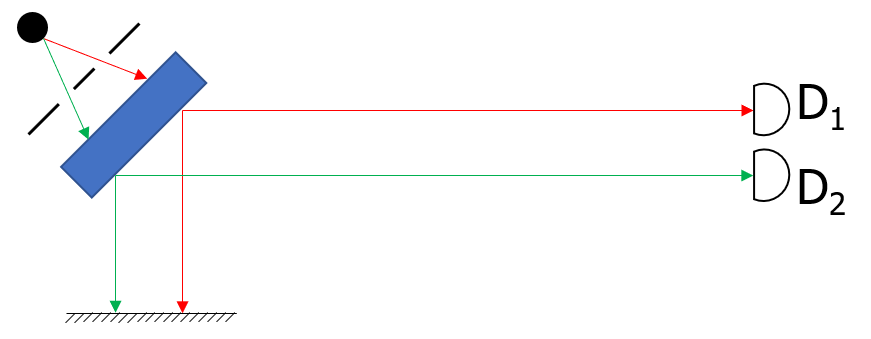}
 \caption{\footnotesize The first quantum eraser experiment.
 The black dot denotes the single-photon source, from which a photon is sent through two slits towards a nonlinear crystal denoted by the blue rectangle. From there one photon is sent towards the screen (down) and one photon is sent towards two detectors which can be used to determine the path it took (right).}
 \label{fig:qunatumeraseronlypaths.png}
\end{figure}

The first experiment (see Figure \ref{fig:qunatumeraseronlypaths.png}) places the detectors $D_1$ and $D_2$ directly in the path of the idler photon.
After completion of the experiment, we split the
dots on the screen created by the signal photons into two groups - one group consisting of the dots corresponding to those pairs whose idler photons were detected in detector $D_1$ and the other group consisting of the dots associated with detector $D_2$. The result of the experiment is that each of these two groups shows a pattern reminiscent of photons moving through a single slit. Neither one of the two groups of dots shows interference. \par

\begin{figure}
 \centering
 \includegraphics[width = 0.7\textwidth]{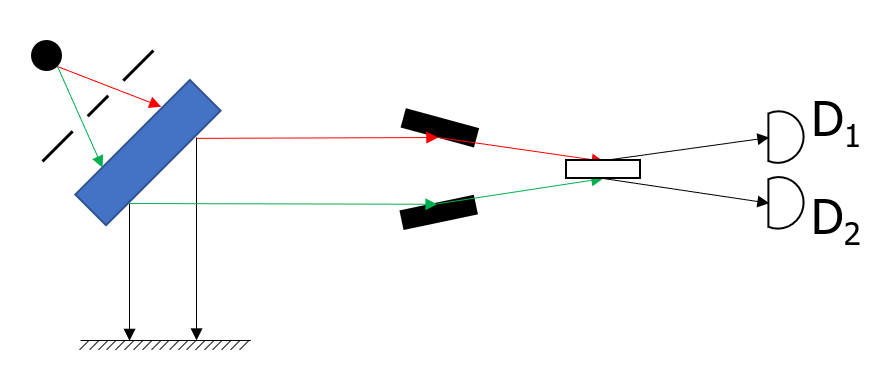}
 \caption{\footnotesize The second quantum eraser experiment.
 The black dot denotes the single-photon source, from which a photon is sent through two slits towards a nonlinear crystal denoted by the blue rectangle. From there, one photon is sent towards the screen (down). The other photon is sent towards the mirror, denoted by black rectangles, and the beamsplitter, denoted by the open rectangles (right). This side ends with the detectors, denoted by the open-half circles. The red and green arrows follow the possible `paths' of the particle. The red and green lines indicate the two possible paths taken by the photon; the last steps are made black to indicate the `erased' path information. The path of the photon should then be read as a superposition between the two.
 \label{fig:qunatumeraserwithbeamsplitter.png}}
\end{figure}

In the second experiment (see Figure \ref{fig:qunatumeraserwithbeamsplitter.png}), the photons are directed towards a beamsplitter before being directed towards the detectors {{$D_1$ and $D_2$.}}
This causes both idler paths to interfere. When we again split the dots created by the signal photons into two groups - one corresponding to the clicks of the detector {{$D_1$}} and one corresponding to the clicks of detector {{$D_2$}}. In this case, both sets show an interference pattern reminiscent of a double-slit experiment. \par
\begin{figure}
 \centering
 \includegraphics[width = 0.9\textwidth]{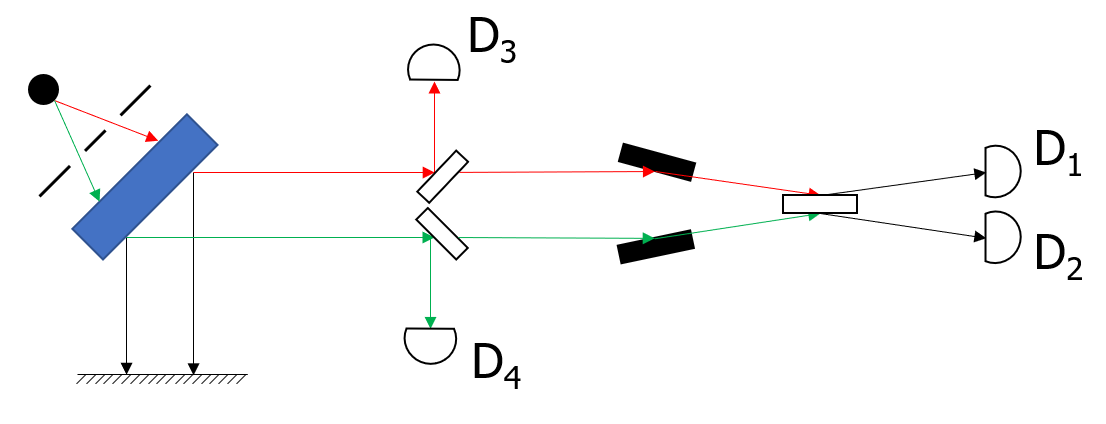}
 \caption{\footnotesize The third quantum eraser experiment. The black dot denotes the single-photon source, from which a photon is sent through two slits towards a nonlinear crystal denoted by the blue rectangle. From there, one photon is sent towards the screen (down). The other photon is sent towards the mirrors, denoted by black rectangles, and the beamsplitters, denoted by the open rectangles (right). This side ends with the detectors, denoted by the open-half circles. The red and green arrows follow the possible `paths' of the particle. The red and green lines indicate the two possible paths taken by the photon; the last steps are made black to indicate the `erased' path information. The path of the photon should then be read as a superposition between the two.
 \label{fig:quantumerasercomplete}}
\end{figure}

In the third experiment, the beamsplitters are set up in such a way that the decision as to which of the above two experiments is performed is the result of pure chance. In contrast to Wheeler's experiment, no random generator is introduced, but the randomness is introduced by replacing the mirrors in the first experiment by beamsplitters (see Figure \ref{fig:quantumerasercomplete}). This means that the (half-)mirrors are set up in such a way that all detectors click with equal probability. When we now group the data of the position on the screen of the signal photon based on which detector clicked that round, we see the patterns as in the previous experiments: when detector $D_1$ {or} $D_2$ clicked, the corresponding data {show} interference, and when detectors $D_3$ {or} $D_4$ clicked, the corresponding data does {not} show interference. \par

\subsection{The argument for retro-causality}

In their discussion of the quantum eraser experiment, both Scully and Dr\"{u}hl \cite{ScullyDruehl} and Kim et.al. \cite{Kim} refrain from any specific interpretative argumentation, although their interest in these set-ups clearly is motivated by an argument of this sort. However, an argument in the spirit of Wheeler is not hard to reconstruct. \par
In this first set-up, the lack of displayed interference is often ``explained'' by saying that detection at the detectors retro-actively reveals which path the photon took. The absence of interference in the joint detection at the screen and at each one of the detectors is said to ``reveal retroactively the particle-like nature of the photon''. Likewise, the displayed interference pattern in the second set-up is often ``explained'' by saying that the interference retro-actively erased the `which path' information. The observed interference in the joint detection at the screen and each one of the detectors is said to ``reveal retroactively the wave-like nature of the photon''. By extending the arm of the idler path, the choice between which aspect of the photon is revealed is crucially delayed. Therefore, the third case can be understood as a retro-active influence on the wave or particle nature of the photon. \par
The apparent improvement of the delayed quantum eraser over Wheeler's original experiment is that even if one takes an agnostic position as to whether the photon ``was'' a particle or a wave, the experiment still appears somewhat puzzling at first sight. As the grouping of data of the signal photon based on the clicked detector of the idler photons exposes previously hidden patterns, we can still group the data of an experimenter placed at the screen in seemingly locally random sets (according to the outcomes of the idler photon) that seem to enforce a certain interpretation (c.f. \cite[p. 483]{Ma2}). That is, for the locally random subsets of signal photon data, we seem to be able to make an interference pattern appear or not appear. \par

\subsection{Analysis of the experiment\label{sec:DQEanalysis}}

\begin{figure*}
\begin{subfigure}{0.45\textwidth}
\includegraphics[width=\linewidth]{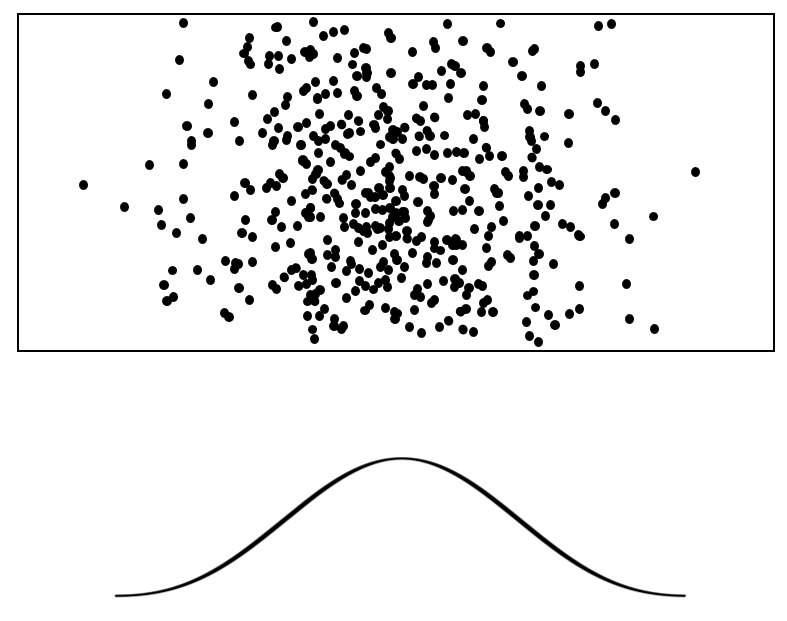}
\caption{\footnotesize Distribution of the dots {\em without} marking their outcomes based on the detector outcome. \label{fig:DQE-1}}
\end{subfigure}
\hfill
\begin{subfigure}{0.45\textwidth}
\includegraphics[width=\linewidth]{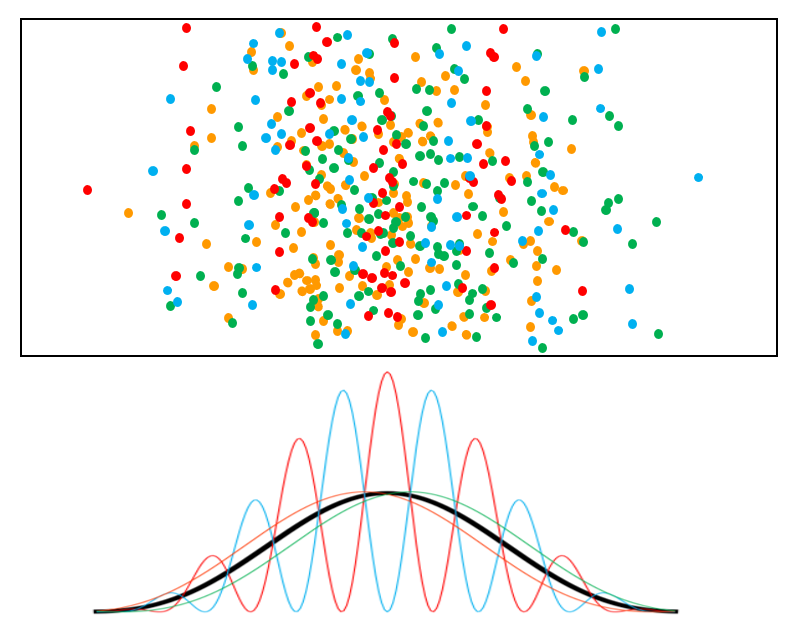}
\caption{\footnotesize Distribution of the dots {\em with} marking their outcomes based on the detector outcome. \label{fig:DQE-3-combined-firstappearance}}
\end{subfigure}

\caption{\footnotesize The key difference between marking and not marking the outcome of the data on the screen based on the detector}\label{fig:DQE-0}
\end{figure*}

The key observation for the analysis of the delayed quantum eraser presented below is summarised by Figure \ref{fig:DQE-0}. That is, only when we group the dots on the screen based on which of the detectors have clicked {\em after} many runs of the experiment, patterns of interference do or do not show \cite{Kim}. In what follows, we build a forward analysis of the experiment based on this principle. We first provide a time evolution of the quantum state describing the three experiments, as in Figures \ref{fig:qunatumeraseronlypaths.png}, \ref{fig:qunatumeraserwithbeamsplitter.png}, and \ref{fig:quantumerasercomplete}, using text-book quantum mechanics only. After this first analysis, we re-cast our initial analysis in the mathematical language of operators as we have done for Wheeler's experiment above. We use this analysis to again show the equivalence of the delayed and non-delayed versions of the experiment. Lastly, we ground this equivalence in the physical principle that space-like separated arms of the experiment necessarily commute.
\par

\subsubsection{Analysis of the first experiment\label{sec:analysisoffirstexpeirmentDQE}}

\begin{figure}[]

\subfloat[The delayed choice quantum eraser with bundles of light instead of rays]{%
 \includegraphics[clip,width=0.95\columnwidth]{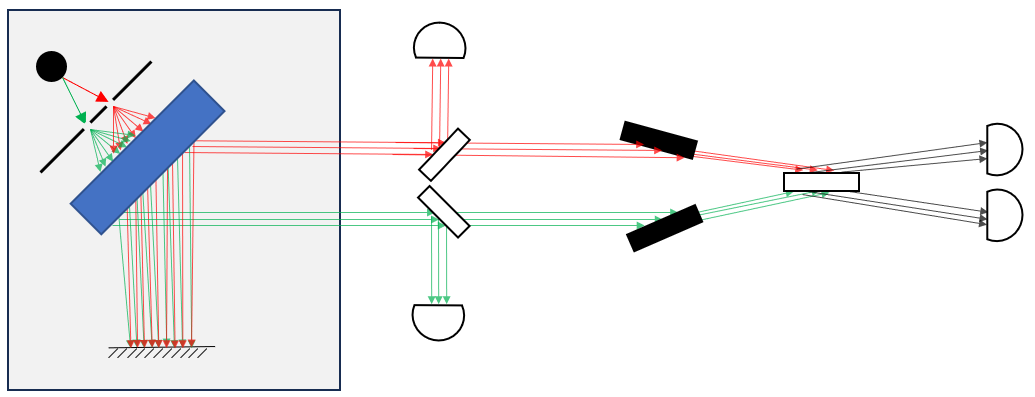}%
 \label{fig:delayederaserwithrays}
}

\bigskip\bigskip
\subfloat[A zoomed in schematic of the delayed quantum eraser containing an indication of the definitions of the angles and outcome positions.]{%
 \includegraphics[clip,width=0.5\columnwidth]{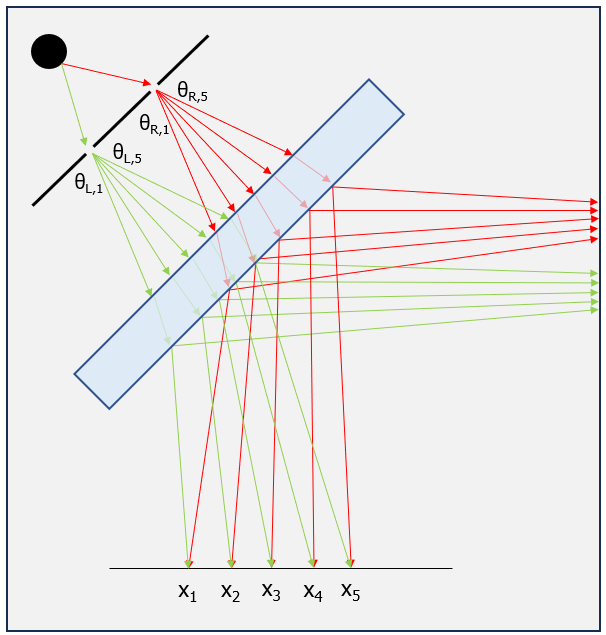}%

}

\caption{\label{fig:delayederaserdefinitionsofangles}}
\end{figure}

In line with our explanation above, we begin by analysing the first experiment (see Figure \ref{fig:qunatumeraseronlypaths.png}). \par
In order to describe the state of the signal photon, we keep track of the angles of the paths of the photon after having passed the respective slits (see Figure \ref{fig:delayederaserdefinitionsofangles}). To good approximation, {we assume that these angles can only take finitely many discrete values
 $\theta_{R,1},\dots,\theta_{R,N}$
(for photons passing through the right slit), respectively
$\theta_{L,1},\dots,\theta_{L,N}$ (for photons passing through the left slit), in the interval $(-\frac12\pi,\frac12\pi)$. These angles are subject to a constraint formulated shortly.}
 The approximation is used to allow for an analysis of the experiment in basic (finite-dimensional) quantum mechanics, which simplifies the (already sufficiently complex) calculation to follow. The original situation can be recovered by taking the limit $N \to \infty$. \par

The state of the system (photon plus screen) after the photon has passed through the double slit can then be described by the superposition
\begin{equation*}
 \ket{1, t = 1} = \frac1{\sqrt{2}}\left(\sum_{n=1}^N p_R(\theta_{R,n}) \ket{R,n}+ \sum_{m=1}^N p_L(\theta_{L,m})\ket{L,m}\right) \ket{\text{ready}}_S.
\end{equation*}
Here, $\ket{R,n}$ and $\ket{L,m}$ describe the state of a photon that passed through the right slit and continued at angle $\theta_{R,n}$, respectively through the left slit and continued at angle $\theta_{L,m}$.
These states are distributed with intensities $p_{R, n} := p_R(\theta_{R,n})$ and $p_{L, m} := p_L(\theta_{L,m})$, respectively, which are simply the normalised intensities after diffraction from the slit. \par
After the photon has passed through the nonlinear crystal, the state of the system can be described as
\begin{equation*}
 \ket{1, t = 2} = \frac1{\sqrt{2}} \left( \sum_{n=1}^N p_{R,n} \ket{R,n}_s\ket{R,n}_i + \sum_{m=1}^N p_{L,m} \ket{L,m}_i\ket{L,m}_s\right)\ket{\text{ready}}_S,
\end{equation*}
where $\ket{R,n}_i\ket{R,n}_s$ and $\ket{L,n}_i\ket{L,n}_s$ refer to the newly created entangled pairs of idler photon and signal photon. \par
 Next, the signal photon will be measured at the screen.
We assume that the angles $\theta_{L,1}, \dots, \theta_{L,N}$ and $\theta_{R,1}, \dots \theta_{R,N}$ are such that, for each $k=1,\dots,N$, the photons corresponding to angles $\theta_{L,k}$ and $\theta_{R,k}$ both arrive at the same position $x_k$ on the the screen. See figure \ref{fig:delayederaserdefinitionsofangles}.

Suppose now that the signal photon is measured to arrive at the screen in partition $x_k$.
As the signal photon gets absorbed, we can model the state as
\begin{equation}
 \label{eq:DQE-1t2nointerference}
 \ket{1, t = 3} = \frac1{\sqrt{2}} \Bigl( \tilde p_{R,k} \ket{R,k}_i + \tilde p_{L,k} \ket{L,k}_i\Bigr)\ket{x_k}_S,
\end{equation}
where $\ket{x_k}_S$ denotes the state of the screen having registered the signal photon in the interval $x_k$.
The numbers $\abs{\tilde p_{R,k}}^2$ and $\abs{\tilde p_{L,k}}^2$ represent the probability density of photons with states $\ket{R,k}$ and $\ket{L,k}$ reaching $x_k$ respectively. These probabilities can be calculated from our initial distribution by simple conditional probabilities, as
\begin{align*}
 & \tilde p_{R,k}(x) = \frac{p_{R,k}}{\sqrt{\abs{p_{R,k}}^2 + \abs{p_{L,k}}^2}}, & \tilde p_{L,k}(x) = \frac{p_{L,k}}{\sqrt{\abs{p_{R,k}}^2 + \abs{p_{L,k}}^2}}.
\end{align*}

\begin{figure*}
\begin{subfigure}{0.45\textwidth}
\includegraphics[width=\linewidth]{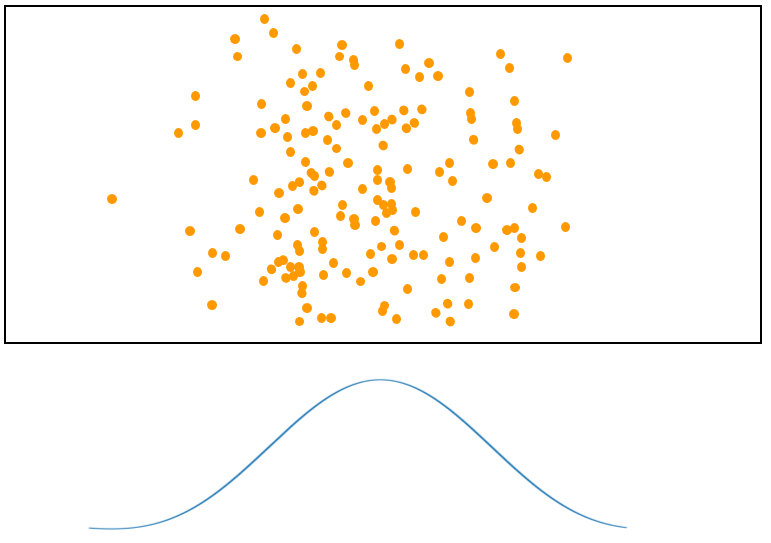}
\caption{\footnotesize Distribution of the dots in group 1 representing the passage through slit $R$ (with slit $L$ closed)}
\end{subfigure}
\hfill 
\begin{subfigure}{0.45\textwidth}
\includegraphics[width=\linewidth]{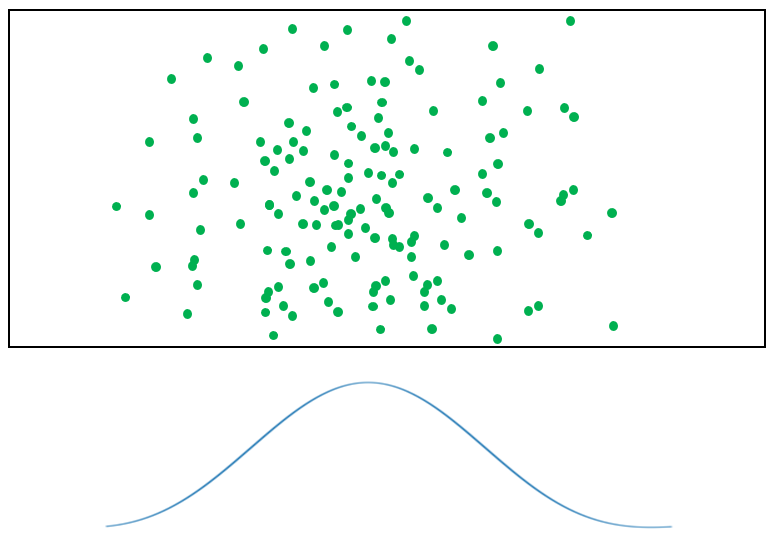}
\caption{\footnotesize Distribution of the dots in group 2 representing the passage through slit $L$ (with slit $R$ closed)}
\end{subfigure}

\caption{\footnotesize Experiment 1: The two groups of dots.}\label{fig:DQE-1LR} 
\end{figure*}

Expression \eqref{eq:DQE-1t2nointerference} clearly shows that {\em the presence of the idler photon prevents interference from taking place}. Upon arrival at position $x_k$, the idler photon is left in a superposition of states whose momentum is perpendicular to the path of the signal photon on its way to
position $x_k$. It is in this sense that the idler photon, after the arrival location of the signal photon has been recorded at the screen, is in a superposition of idler states carrying `which path' information.
In the limit for $N\to \infty$, this corresponds to the idler photon being in a superposition of two states with well-defined angles $\vartheta_{R,x}$ and $\vartheta_{L,x}$ relative to the crystal. These angles are uniquely determined by the position $x$ on the screen.
Effectively, the distribution on the screen will now be described by the normalised addition
of the distributions corresponding to the two `which path' scenarios corresponding to passage through slit $L$ (with slit $R$ closed) and $R$ (with slit $L$ closed) (see Figure \ref{fig:DQE-1LR}).

The phenomenon that the interference pattern disappears in the presence of idler photons,
regardless of whether or not one actually measures them,
was described by Zeilinger \cite{Zeilinger} in an equivalent set-up
as follows (additions in brackets by the present authors):
\begin{quote}
``(\dots) whenever particle 1 [the idler photon] is found in beam $a$, particle 2 (the signal photon) is found in beam $b$ and whenever
particle 1 is found in beam $a'$, particle 2 is found in beam $b'$. The quantum state is
$$ \ket{\psi} = \frac1{\sqrt{2}}(\ket{a}_1\ket{b}_2+\ket{a'}_1\ket{b'}_2).$$
Will we now observe an interference pattern for particle 1 behind its double slit? The answer has
again to be negative because by simply placing detectors in the beams $b$ and $b'$ of particle $2$ (the idler photon) we can determine which path particle 1 took. Formally speaking, the states $\ket{a}_1$ and $\ket{a'}_1$ again cannot be coherently superposed because they are entangled with the two orthogonal states $\ket{b}_1$ and $\ket{b'}_1$.

Obviously, the interference pattern can be obtained if one applies a so-called quantum eraser which completely
erases the path information carried by particle 2. That is, one has to measure particle 2 in such a way that
it is not possible, even in principle, to know from the measurement which path it took, $a'$ or $b'$.''
\end{quote}
The above analysis, which is similar in spirit to the one presented for the `double double-slit experiment in \cite{double-double}, makes this precise.

\subsubsection{Analysis of the second experiment}

In the second experiment (of Figure \ref{fig:qunatumeraserwithbeamsplitter.png}), the `which path' information contained in the angles $\vartheta_{R,x}$ and $\vartheta_{L,x}$ of the idler photons is erased through the introduction of a beamsplitter. Naively, one would expect that
therefore, in this configuration, an interference pattern should build up after all, and that contextual information is required: the photon, in order to be able to decide which probability distribution to ``use'', must ``know'' in advance the experimental context in which it finds itself. As the ensuing analysis will show, no such information is needed and, in fact, no interference will build up.

Let us partition the arrival at the screen into two groups, $G_1$ and $G_2$; the first consists of those arrivals whose idler partners made detector $D_1$ click, and the second of those dots whose idler partners made detector $D_2$ click.
To include the two-detector system comprising $D_1$ and $D_2$ into the considerations,
we introduce a ready-to-measure state for this system. Thus we replace the
state $\ket{1, t = 3}$ of equation \eqref{eq:DQE-1t2nointerference} by
\begin{align*} \ket{2, t = 3} & = \frac1{\sqrt{2}}\Bigl( \tilde p_{R,k}\ket{R,k}_i + \tilde p_{L,k}\ket{L,k}_i\Bigr)\ket{x_k}_S\ket{\text{ready}}_D,
\intertext{
and similarly for $\ket{1, t = 2}$.
Having passed the beamsplitter, before reaching the detector system the state can be described as
 }
\ket{2, t = 4} & =
\frac12\Bigl((\tilde p_{R,k}\ket{R,k,1}_i + \tilde p_{R,k}\ket{R,k,2}_i)
\\ & \qquad + (\tilde p_{L,k}\ket{L,k,1}_i - \tilde p_{L,k}\ket{L,k,2}_i)\Bigr)\ket{x_k}_S\ket{\text{ready}}_D,
\intertext{
with newly labelled idler states indicating which detectors lie in their paths.
Upon arrival at the detectors the idler photons are absorbed, leaving the system in the (as yet unobserved) state
 }
\ket{2, t = 5} & =
\frac12 \Bigl( (\tilde p_{R,k}\ket{1}_D + \tilde p_{R,k}\ket{2}_D)
 +(\tilde p_{L,k}\ket{1}_D - \tilde p_{L,k}\ket{2}_D)\Bigr)\ket{x_k}_S,
\\ & = \frac12 \Bigl((\tilde p_{R,k} + \tilde p_{L,k})\ket{1}_D
 + (\tilde p_{R,k}- \tilde p_{L,k})\ket{2}_D\Bigr)\ket{x_k}_S,
\end{align*}
where $\ket{1}_D$ and $\ket{2}_D$ describe the states of the detector system in which $D_1$ respectively $D_2$ has registered a photon.

\begin{figure*}
\begin{subfigure}{0.45\textwidth}
\includegraphics[width=\linewidth]{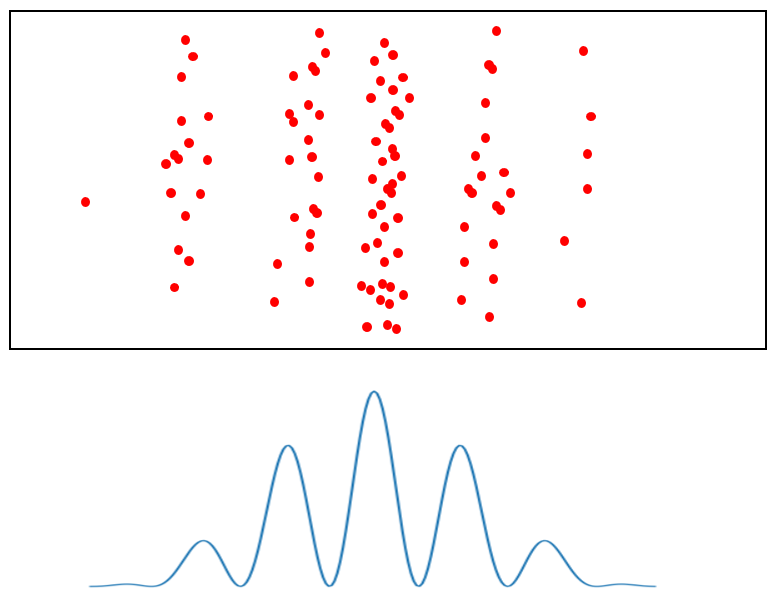}
\caption{\footnotesize Distribution of the dots in group 1}
\end{subfigure}
\hfill 
\begin{subfigure}{0.45\textwidth}
\includegraphics[width=\linewidth]{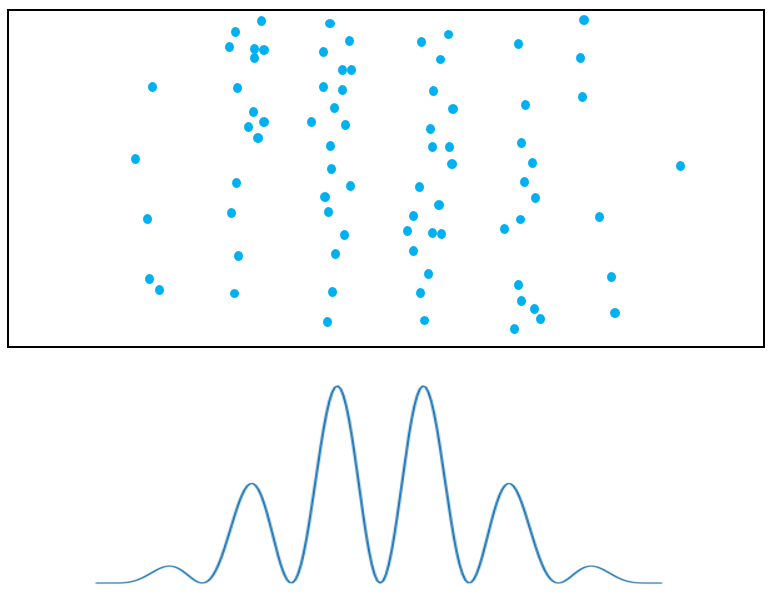}
\caption{\footnotesize Distribution of the dots in group 2}
\end{subfigure}

\caption{\footnotesize Experiment 2: The two groups of dots}\label{fig:DQE-2} 
\end{figure*}

In order to arrive at an expression for the distribution of the $G_1$-arrivals we write
$$ \ket{1}_D = \ket{\text{click}}_{D_1} \ket{\text{no click}}_{D_2}, \qquad
 \ket{2}_D = \ket{\text{no click}}_{D_1} \ket{\text{click}}_{D_2}$$
 and trace out detector $D_2$. This results in the reduced state (density matrix)
\begin{align}
\label{eq:DQEexp2t=5D2tracedout}
\ket{2, t = 5}\bra{2, t = 5}_{D_2 \, \text {traced out}} & = \frac{1}{4}\Bigl( \abs{\tilde p_{R,k}}^2 + 2\ \Re\big(\tilde p_{R,k}\overline{\tilde p_{L,k}}\big) \\
& \qquad + \abs{\tilde p_{L,k}}^2\Bigr)\ket{\text{click}}_{D_1}\ket{x_k}_S\bra{x_k}_S\bra{\text{click}}_{D_1} \nonumber \\
& \phantom{=} + \frac{1}{4}\Bigl( \abs{\tilde p_{R,k}}^2 - 2\ \Re\big(\tilde p_{R,k}\overline{\tilde p_{L,k}}\big) \nonumber \\
& \qquad + \abs{\tilde p_{L,k}}^2\Bigr)\ket{\text{no click}}_{D_1}\ket{x_k}_S\bra{x_k}_S\bra{\text{no click}}_{D_1} \nonumber
\intertext{
On the basis of this state, we expect that the distribution of the $G_1$-arrivals, that is, the joint detection of the events $\{$arrival at $x_k$ and $D_1$ clicked$\}$ shows the same interference as in the Young double slit experiment.
Likewise, tracing out $D_1$ results in the reduced state}
\label{eq:DQEexp2t=5D1tracedout}
\ket{2, t = 5}\bra{2, t = 5}_{D_1 \, \text {traced out}} & = \frac{1}{4}\Bigl( \abs{\tilde p_{R,k}}^2 + 2\ \Re\big(\tilde p_{R,k}\overline{\tilde p_{L,k}}\big) \\
& \qquad + \abs{\tilde p_{L,k}}^2\Bigr)\ket{\text{no click}}_{D_2}\ket{x_k}_S\bra{x_k}_S\bra{\text{no click}}_{D_2} \nonumber \\
& \phantom{=} + \frac{1}{4}\Bigl( \abs{\tilde p_{R,k}}^2 - 2\ \Re\big(\tilde p_{R,k}\overline{\tilde p_{L,k}}\big) \nonumber \\
& \qquad + \abs{\tilde p_{L,k}}^2\Bigr)\ket{\text{click}}_{D_2}\ket{x_k}_S\bra{x_k}_S\bra{\text{click}}_{D_2} \nonumber
\end{align}
From this we infer that also the distribution of the $G_2$-arrivals shows the interference of the Young double slit experiment,
but with a shift in the $x_k$-variable caused by the phase shift over $\pi$ due to the presence of the minus sign (see Figure \ref{fig:DQE-2}).
\newline\indent
The distribution of all arrivals, comprising both the $G_1$-arrivals and the $G_2$-arrivals,
is obtained by tracing out both detectors, which results in the reduced state
\begin{equation}
\label{eq:DQEexp2t=5allDtracedout}
\ket{2, t = 5}\bra{2, t = 5}_{D \, \text{traced out}} = \ket{x_k}_S\bra{x_k}_S.
\end{equation}
We see that in this case no interference is built up. \par
On the basis of the above calculations, and in line with the heuristic argument of \cite{Hossenfelder}, we conclude that if the data is grouped in subsets based on whether detector 1 or detector 2 clicked, \eqref{eq:DQEexp2t=5D2tracedout} and \eqref{eq:DQEexp2t=5D1tracedout} predict the emergence of the interference patterns as depicted in Figure \ref{fig:DQE-2}. Furthermore, if this data is not grouped on the basis of which detector clicked, then no interference is detected and its outcome is equal to the distribution given in Figure \ref{fig:DQE-1}.

\subsubsection{Analysis of the third experiment}

\begin{figure*}[ht]
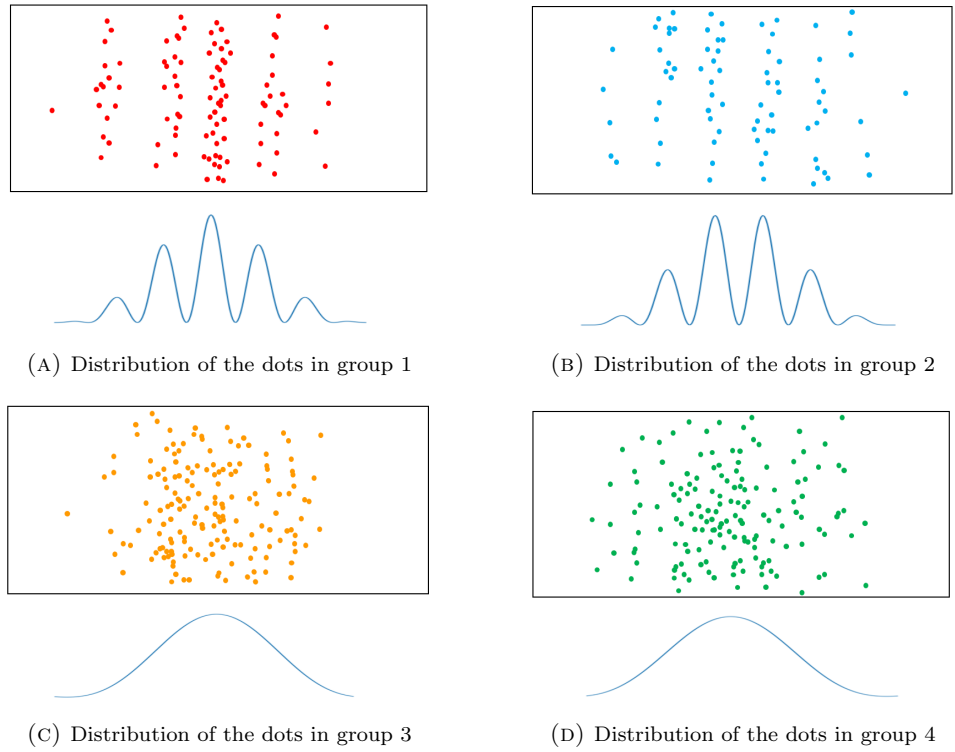
 

\begin{subfigure}{0.45\textwidth}
\includegraphics[width=\linewidth]{pictures/outcomedetector3bob1screenwithcolor.png}
\caption{\footnotesize Distribution of the dots in group 1}
\end{subfigure}
\hfill 
\begin{subfigure}{0.45\textwidth}
\includegraphics[width=\linewidth]{pictures/outcomedetector4bob2screenwithcolor.png}
\caption{\footnotesize Distribution of the dots in group 2}
\end{subfigure}

\bigskip 
\begin{subfigure}{0.45\textwidth}
\includegraphics[width=\linewidth]{pictures/outcomedetector1alice1screenwithcolor.png}
\caption{\footnotesize Distribution of the dots in group 3}
\end{subfigure}
\hfill 
\begin{subfigure}{0.45\textwidth}
\includegraphics[width=\linewidth]{pictures/outcomedetector2alice2screenwithcolor.png}
\caption{\footnotesize Distribution of the dots in group 4}
\end{subfigure}
\caption{\footnotesize Experiment 3: The four groups of dots}
\label{fig:DQE-3} 
\end{figure*}

\begin{figure}[ht]
 \centering
 \includegraphics[scale=0.5]{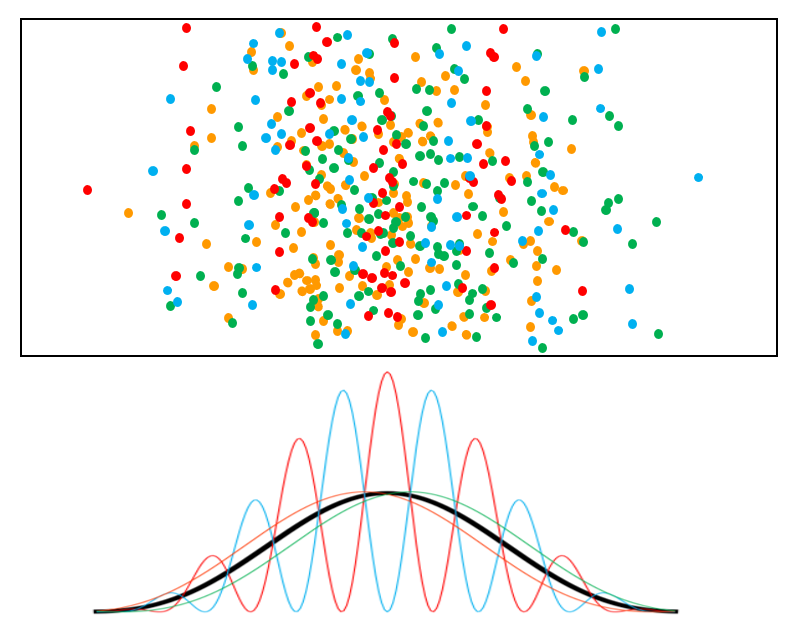}
 \caption{\footnotesize Experiment 3: All dots combined}
 \label{fig:DQE-3-combined}
\end{figure}

In this experiment, the previous two experiments are combined by introducing a second beamsplitter in such a way that the idler photons reflected by the beamsplitter to one of the detectors $D_1$ or $D_2$ will {\em not} reveal `which path' information, since
the `which path' information of the idler photons passing through the beamsplitter will be erased by the next beamsplitter in their path. Clicks of the detectors $D_3$ and $D_4$ {do} reveal `which path' information. The reader will have no difficulty working out the formulas describing the succession of states in this scenario; the reasoning follows the same patterns as in the preceding two cases.

This time we can group the dots on the screen into four groups, corresponding to which of the detectors $D_{1-4}$ clicked.
The results for these groups are depicted in Figure \ref{fig:DQE-3}. In line with the first experiment, both the $H_1$-arrivals corresponding to registrations at detector $D_1$ and the $H_2$-arrivals corresponding to registrations at $D_2$ show interference (as they reveal no `which-path' information), but the combined arrivals of $H_1$ and $H_2$ add up to a pattern without interference.
In line with the second experiment, the $H_3$-arrivals and the $H_4$-arrivals corresponding to registrations at $D_3$ and $D_4$, respectively, show no interference (as they reveal `which path' information).

\subsubsection{Mathematical analysis and equivalence with the non-delayed experiment}\label{sec:DQEmathequivalence}

This section will give a mathematical analysis of the delayed quantum eraser
similar to that presented in Section \ref{sec:wheelermathequivalence}, for the scenario with delayed eraser and a version of it without delay. Again these scenarios turn out the produce the same final state.

Throughout the subsequent analysis, we fix a positive integer $N$; only after performing all calculations that we interpret the results in passing to the limit $N\to\infty$. To good approximation, we assume that the photon, when passing the double slit, chooses between $N$ fixed angles $\theta_1,\dots,\theta_N\in (-\frac12\pi,\frac12\pi)$.
The state of the photon, once it passed through the double slit,
can be modelled by an element of the Hilbert space $\C^2\otimes \C^N$. In this representation, the standard basis vectors $\ket{R,n} := \ket{R}\ket{n}$ and $\ket{L,n} = \ket{L}\ket{n}$ describe the state of a photon passing through the right, respectively left, slit and emanating from it at angle $\theta_n$.

The nonlinear crystal is modelled by a $(2N\times 4N)$-matrix $C$ acting from $\C^2\otimes \C^N$ to $(\C^2\otimes \C^N)\otimes (\C^2\otimes \C^N)$ with action
\begin{align*}
C: \begin{cases}
\ket{R}\ket{n} & \mapsto \ \ \ket{R_s}\ket{n}\otimes \ket{R_i}\ket{n}, \\
\ket{L}\ket{n} & \mapsto \ \ \ket{L_s}\ket{n}\otimes \ket{L_i}\ket{n},
\end{cases}\end{align*}
where the indices $s$ and $i$ on the right-hand side are nothing but a notational device to keep track of the signal and idler photons, respectively.
We ignore the fact that the idler and signal photons have halved frequencies; this plays no role in the present qualitative analysis.

The action of the screen is modelled by any $(4N\times 2N)$-matrix $S$ acting from $(\C^2\otimes\C^N)\otimes (\C^2\otimes \C^N)$
to $\C^2\otimes \C^N$ with action
\begin{align*}
S: \begin{cases}\displaystyle \ket{R_s}\ket{n}\otimes \ket{R_i}\ket{n} & \mapsto \ \ \displaystyle \frac1{\sqrt{N}}\sum_{n=1}^N \tilde p_{R,n}\ket{R_i}\ket{n}, \\
\ket{L_s}\ket{n}\otimes \ket{L_i}\ket{n} & \mapsto \ \ \displaystyle \frac1{\sqrt{N}}\sum_{n=1}^N \tilde p_{L,n}\ket{L_i}\ket{n},
\end{cases}
\end{align*}
The action of the beamsplitters and mirrors may be lifted to the liner operators $H\otimes I$ and $X\otimes I$ on $\C^2\otimes \C^N$.
The configuration of two beamsplitters behind the non-linear crystal (see Figure \ref{fig:quantumerasercomplete}) acts
as one beamsplitter, provided we interpret $R$-photons as `up' and $L$-photons as `down', and interpret the photons deflected towards detectors $D_1$ and $D_2$ as `down'.

By following the sequence of elements {encountered by} the photon, we can determine the operator representing the delayed quantum eraser. We see that the photon, after going through the first slits, encounters the crystal $C$, the Screen $S$, then the first beamsplitter $(H\otimes I)$, the mirrors $(X \otimes I)$ and, lastly, the second beamsplitter $(H\otimes I)$. The delayed choice experiment may therefore be represented by the composition
\begin{align}%
\label{eq:DQE-math-delay}
A = (H\otimes I)\circ (X\otimes I)\circ (H\otimes I) \circ S \circ C = (Y \otimes I)\circ S\circ C,
\end{align}
where $Y = \bigl(\begin{smallmatrix} 1 \ \phantom{-}0 \\ 0 \ -1\end{smallmatrix}\bigr)$.

As in the case of Wheeler's delayed choice experiment, we may rearrange the order of steps in such a way that
no delayed choice takes place. This will be done by making the distance from the crystal to the screen long, and the distance to the detector short. Here, the beamsplitters are interpreted as acting on
$(\C^2\otimes \C^N)\otimes(\C^2\otimes \C^N)$ as $(I\otimes I)\otimes (H\otimes I)$
since the beamsplitter acts trivially on the signal photons; the mirrors can be represented in the same way. In this case, after the photon has passed the crystal $C$, the idler photon will pass the first beamsplitter ($(I\otimes I)\otimes (H\otimes I)$), the mirrors ($(I\otimes I)\otimes (X\otimes I)$) and the second beamsplitter ($(I\otimes I)\otimes (H\otimes I)$). Lastly, the signal photon will encounter the screen $S$.
This version of the experiment may now represented by the composition
\begin{align}%
\label{eq:DQE-math-no-delay}
A' & = S \circ ((I\otimes I)\otimes (H\otimes I))\circ((I\otimes I)\otimes (X\otimes I))\circ ((I\otimes I)\otimes (H\otimes I))\circ C
\\ \nonumber & = S\circ ((I\otimes I)\otimes (Y\otimes I))\circ C
\\ \nonumber & = (Y\otimes I)\circ S\circ C,
\end{align}
where the last equality is immediate from the definition of $S$. Combining equations \eqref{eq:DQE-math-delay} and \eqref{eq:DQE-math-no-delay}, we see again that $A = A'$. Therefore, as in the case of the delayed choice experiment, we conclude that {\em the scenarios with and without delayed choice result in the same final state}.

\subsubsection{The operational equivalence of the delayed and non-delayed experiment in relation to special relativity.}
\begin{figure}[H]
\centering
\includegraphics[width = 1\textwidth]{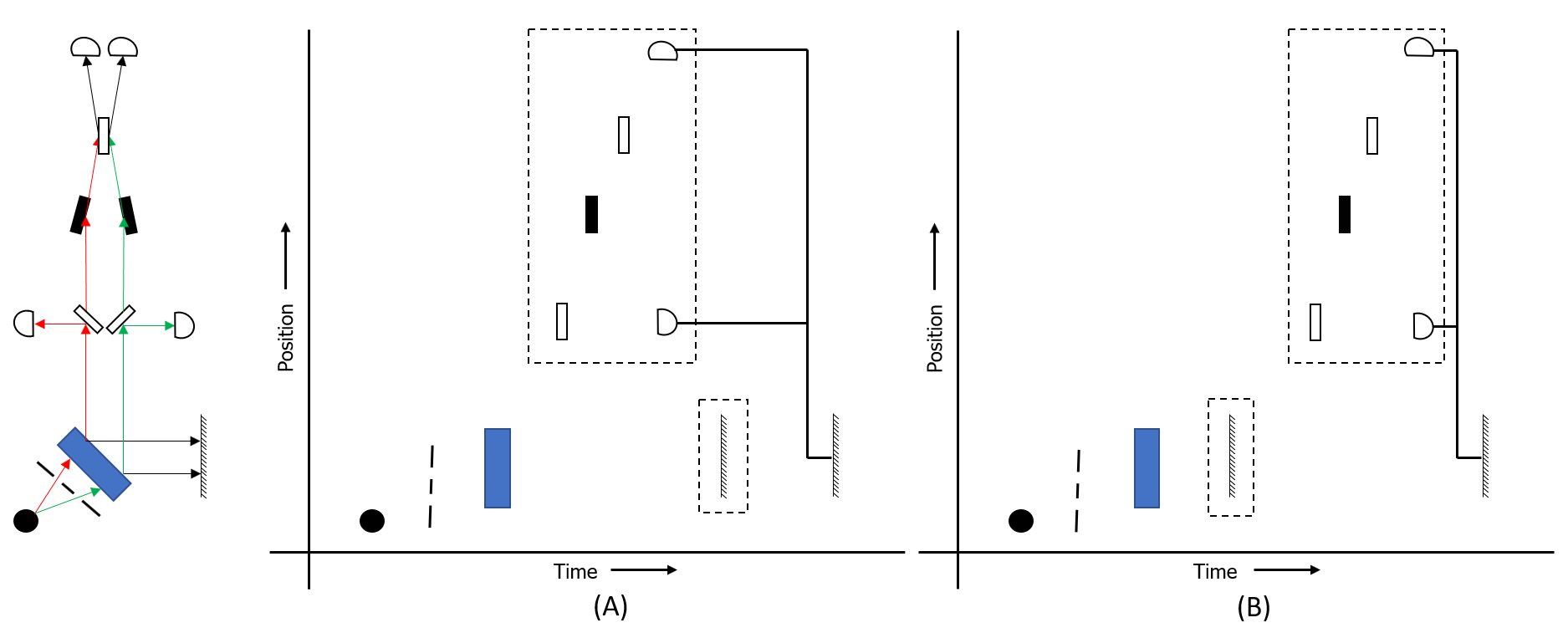}
\caption{A schematic depiction of the space-time diagram of a non-delayed quantum eraser. The left side of the figure denotes the standard set-up of the delayed quantum eraser, as in Figure \ref{fig:quantumerasercomplete}. The right side of the figure places this set-up in a time-ordered sequence. The black dot denotes the single-photon source, from where a photon is sent through two slits towards a nonlinear crystal denoted by the blue rectangle. From there, one photon is sent towards some gray medium, slowing its trajectory. From there it moves towards the screen, after the other photon is registered by the detectors. This other photon is first sent towards the mirrors, denoted by black rectangles, and the beamsplitters, denoted by the open rectangles. This side ends with the detectors, denoted by the open-half circles. The lines indicate the causal relation between the detectors, which are used to colour the eventual outcome on the screen. The dotted lines are used to indicate that the two enclosed regions are space-like separated. Schematic (A) shows the space-time diagram of the set-up when the detection of the idler photon is not delayed, (B) shows the space-time diagram of the delayed set-up.}
 \label{fig:delayederaserspacetime}
\end{figure}

To conclude, let us analyse the space-time structure of the delayed quantum eraser. In contrast to the Wheeler delayed choice experiment, in the delayed quantum eraser the two arms of the experiment do not seem to be causally connected. This assessment is not correct, however. In order to distinguish the patterns related to each detector on the screen, the output of the detectors needs to be combined with the screen data after the measurements of both arms of the experiment have been completed. This subtle, but crucial, point brings this case very close to the previous case. Figure \ref{fig:delayederaserspacetime} shows how the delayed quantum eraser can be split up into two space-like separated parts, which
commute. The insight from Figure \ref{fig:delayederaserspacetime} thus physically grounds the equality $A=A'$ in the language of special relativity. Furthermore, in our view a comparison between Figures \ref{fig:delayederaserspacetime} and \ref{fig:wheeler_space_time_complete} shows the crucial similarity between the two experiments from a relativistic point of view.

\section{Discussion and conclusions}\label{sec:disc-concl}
The controversies surrounding delayed choice experiments can only be truly resolved by a `forward' understanding of the experiment, rather than by a `backward' analysis. As we have demonstrated, such a forward analysis is indeed possible, rendering Wheeler's delayed choice experiment and the delayed quantum eraser no more puzzling than anything else involving superposition and/or entanglement. At any given moment during or after the experiments, any agent involved in the experiment can fully explain the data collected he/she has access to at that moment. \par
At no point in the experiments, information from the future or contextual information is needed to explain or predict what happens next. To paraphrase Wheeler, in his thought experiment the photon at the first beamsplitter does not need to ``know'' about the full configuration of the experiment to ``decide'' between wave-like or particle-like behaviour. In fact, questions such as whether the photon ``was'' a wave or particle at the various stages of the experiment -- the centerpiece in arguments purporting to demonstrate retro-causation -- are completely meaningless from an operational point of view and can only lead to pseudo-problems.\par 
While the problems with delayed-choice experiments are often connected to a realist interpretation of the wave-function (as in Ma et al. \cite{Ma2}), a realist following the analysis presented in this paper will not encounter any problems with delayed choice experiments. Rather, the root of the problem seems to be in the use of physical concepts such as ‘wave-particle duality’ or ‘which path information’ as explanatory devices rather than descriptive tools providing heuristic pictures. In the case of Wheeler’s gedanken-experiment, the which-path question becomes meaningful after the random generator has provided an outcome; in the case of the delayed eraser, the which-path question becomes meaningful after the idler photon has passed the beamsplitters. Only after these crucial steps the presence or absence of interference can be argued for on the basis of ‘which path information’. Such an explanation, therefore, involves a certain degree of ‘backwards’ reasoning. It is this `backwards' reasoning that stands at the core of the problems surrounding delayed choice experiments, as it leads to the questioning of past states without a present record. 
Wheeler himself wrote \cite{Wheeler2} 
\begin{quote} Does this result [the delayed-choice experiment] mean that present choice influences past dynamics,
in contravention of every formulation of causality? Or
does it mean, calculate pedantically and don't ask questions?
Neither; the lesson presents itself rather as this, that
the past has no existence except as it is recorded in the
present.\footnote{The existence of records in the present from the past as the only valid way to make inferences about past states as been recently highlighted in our analysis \cite{Waaijer-vanNeerven-FoundPhys} of the celebrated Frauchiger-Renner paradox \cite{F-R}. This analysis reveals that the very existence of records of past measurements by the two Wigner's Friends in the scenario discussed by Frauchiger and Renner can lead to different predictions of certain measurement outcomes in the present.}
\end{quote}
In the experiments discussed in the present paper we have shown that this `backwards' reasoning and the related questioning of past states by the use of heuristic physical concepts can be avoided completely if one adheres to a strictly mathematical analysis. Paraphrasing Zeilinger's words from \cite{Zeilinger}, in the context of the delayed quantum eraser, the disappearance of interference should not be explained contextually on the basis that ``which-path information is still available'', but on the basis of a step-by-step forward analysis of the type presented here.

\medskip\noindent{\bf Data availability statement} -- There are no data attached to this paper.

\medskip\noindent{\bf Conflict of interests statement} -- There is no conflict of interests, and no third parties are involved in this research.

\end{document}